\documentclass[prd,nofootinbib,preprint,superscriptaddress,twocolumn,10pt]{revtex4-1}
\pdfoutput=1
\usepackage{amsmath,amssymb}
\usepackage{epsfig}
\usepackage{graphicx}
\usepackage[normalem]{ulem}
\usepackage[usenames,dvipsnames]{color}
\usepackage{subfigure}
\usepackage{slashed}
\usepackage[colorlinks,citecolor=blue]{hyperref}
\usepackage{color}

\usepackage{multirow}

\usepackage{xcolor}
\usepackage{soul}

\begin{document}
\title{Probing \textit{Miracle-less WIMP} Dark Matter via Gravitational Waves Spectral Shapes}

\author{Debasish Borah}
\email{dborah@iitg.ac.in}
\affiliation{Department of Physics, Indian Institute of Technology Guwahati, Assam 781039, India}
\author{Suruj Jyoti Das }
\email{suruj@iitg.ac.in}
\affiliation{Department of Physics, Indian Institute of Technology Guwahati, Assam 781039, India}
\author{Abhijit Kumar Saha}
\email{psaks2484@iacs.res.in}
\affiliation{School of Physical Sciences, Indian Association for the Cultivation of Science, 2A $\&$ 2B Raja S.C. Mullick Road, Kolkata 700 032, India}
\author{Rome Samanta}
\email{samanta@fzu.cz}
\affiliation{CEICO, Institute of Physics of the Czech Academy of Sciences, Na Slovance 1999/2, 182 21 Prague 8, Czech Republic}

\begin{abstract}
We propose a novel probe of weakly interacting massive particle (WIMP) dark matter (DM) candidates of a wide mass range which fall short of the required annihilation rates to satisfy correct thermal relic abundance, dubbed as \textit{Miracle-less WIMP}. If the DM interactions are mediated by an Abelian gauge boson like B-L, its annihilation rates typically remain smaller than the WIMP ballpark for very high scale B-L symmetry breaking, leading to overproduction. The thermally overproduced relic is brought within observed limits via late entropy dilution from one of the three right handed neutrinos (RHN) present for keeping the model anomaly free and generating light neutrino masses. Such late entropy injection leads to peculiar spectral shapes of gravitational waves (GW) generated by cosmic strings, formed as a result of B-L symmetry breaking. We find interesting correlation between DM mass and turning frequency of the GW spectrum with the latter being within reach of future experiments. The two other RHNs play major role in generating light neutrino masses and baryon asymmetry of the universe via leptogenesis. Successful leptogenesis with Miracle-less WIMP together restrict the turning frequencies to lie within the sensitivity limits of near future GW experiments.



\end{abstract}

\maketitle 

\noindent {\bf Introduction:} Weakly interacting massive particle (WIMP) has been the most widely studied dark matter (DM) candidate (see \cite{Arcadi:2017kky} for a recent review).
However, null results at different WIMP search experiments have motivated the particle physics community to pursue other alternatives like feebly interacting massive particle (FIMP) \cite{Hall:2009bx,Bernal:2017kxu}. In spite of being a viable alternative, FIMP models are often difficult to probe due to tiny couplings of DM with the standard model (SM) bath. Here we consider a new scenario with a wide range of DM masses where DM-SM interaction rates fall short of the required WIMP DM criteria, but large enough to produce it in thermal equilibrium. While typical WIMP DM mass is restricted to be within a few GeV \cite{Lee:1977ua} to few hundred TeV \cite{Griest:1989wd}, the class of DM we study here, dubbed as \textit{Miracle-less WIMP}\footnote{This term was coined recently in \cite{Arakawa:2021vih} in the context of scalar triplet DM abandoning the requirement of relic generation via standard freeze-out (FO) mechanism.}, can have a much wider range of masses. Due to intermediate annihilation rates, such DM gets thermally overproduced and additional mechanism should be in place to bring it within observed limits. Moreover, alternative search strategies need to be adopted for such DM which may not show up in conventional DM searches looking for WIMP DM.

In this letter, we propose a novel way of testing this special class of DM at future gravitational wave (GW) experiments. The DM stability and interactions are taken care of by an Abelian gauge symmetry\footnote{Choice of additional $U(1)$ symmetry is motivated from combining the \emph{Miracle-less WIMP} with the cosmic string formation. The $U(1)_{B-L}$, in particular, naturally accommodates heavy neutral fermions from anomaly cancellation requirements, playing crucial role in generating correct DM relic, neutrino mass and baryon asymmetry via leptogenesis.} which gets broken spontaneously, leading to the formation of cosmic strings (CS) \cite{Kibble:1976sj, Nielsen:1973cs}. These CS can generate stochastic GW with a characteristic spectrum which can be within the reach of near future GW detectors if the scale of symmetry breaking is sufficiently high \cite{Vilenkin:1981bx,Turok:1984cn}. Such a high scale symmetry breaking, leading to a superheavy $Z'$ gauge boson ensures that DM-SM interactions remain in the \textit{Miracle-less WIMP} ballpark. The symmetry breaking scale also determines the mass scales of other heavier particles which serve non-trivial role in generating correct DM abundance as well as the baryon asymmetry of the universe. In particular, entropy production from a heavy long-lived particle can bring down the relic of thermally overproduced DM within observed limits. Interestingly, such late entropy dilution also leads to unique spectral breaks in the GW spectrum generated by CS network. While GW spectral shapes in non-standard cosmology like early matter domination have been studied in earlier works \cite{Cui:2017ufi,Cui:2018rwi,Gouttenoire:2019kij}, we provide a realistic particle physics scenario where such non-standard epoch arises naturally to play non-trivial role in generating correct DM relic abundance. We 
obtain unique correlations between spectral breaks or turning frequencies and DM masses, having the potential of being verified in near future GW experiments. Such intriguing connection makes our present analysis different from the earlier attempts related to GW probe of light DM models \cite{Yuan:2021ebu, Tsukada:2020lgt, Chatrchyan:2020pzh}, superheavy DM models \cite{Bian:2021vmi, Samanta:2021mdm} or dark sector models in general \cite{Schwaller:2015tja} (see \cite{Bertone:2019irm} for review of such possibilities). Additionally, the same framework can also explain the observed baryon asymmetry of the universe via leptogenesis surviving the late entropy dilution. The simultaneous requirement of successful leptogenesis and Miracle-less WIMP DM relic also keep the turning frequencies within the reach of near future GW experiments. \\


\begin{figure}[t]
\includegraphics[scale=0.35]{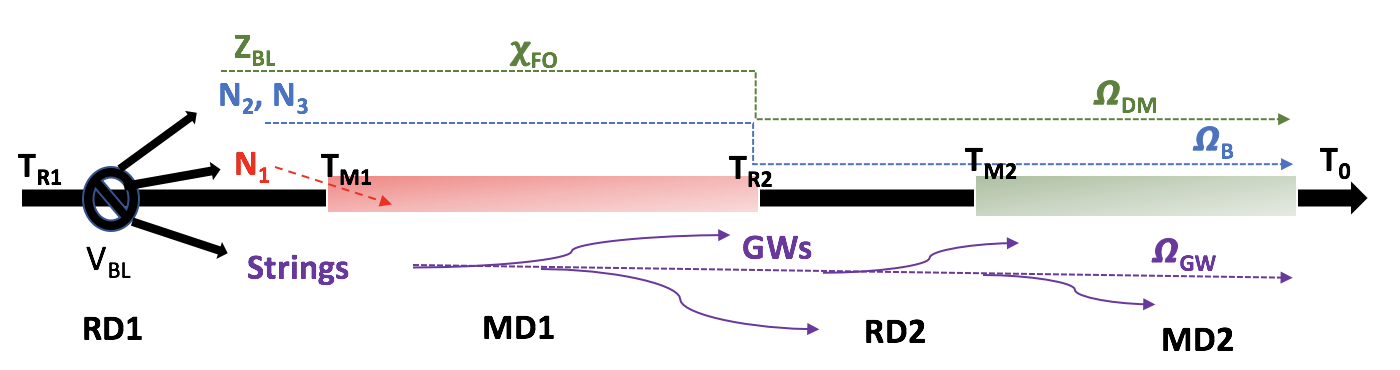} 
\caption{Schematic of the important phases in evolution of the universe from $T_{R_1}$ to $T_0$.} 
\label{fig:Sketch} 
\end{figure}

\noindent {\bf The Framework:} The framework can be based on any Abelian extension of the SM which gives rise to DM stability, interactions in addition to the formation of cosmic strings due to spontaneous breaking. Here we consider the example of gauged $U(1)_{B-L}$ model \cite{Davidson:1978pm, Mohapatra:1980qe, Marshak:1979fm, Masiero:1982fi, Mohapatra:1982xz, Buchmuller:1991ce}. In addition to the SM content, the minimal version of this model contains three right handed neutrinos (RHN) ($N_i$) which take part in type I seesaw origin of light neutrino masses and one complex singlet scalar ($\Phi$) all of which are singlet under the SM gauge symmetry. 
An additional vector like fermion $(\chi)$ with suitable $B-L$ charge $q_\chi$ is considered to be a stable DM candidate. The DM interacts with SM via $Z_{\rm BL}$ mediated processes due to the interaction $\overline{\chi} \gamma^{\mu} {Z_{\rm BL}}_{\mu} \chi$. After spontaneous breaking of both $B-L$ and electroweak symmetries, the $B-L$ gauge boson and RHNs acquire masses as $M_{Z_{\rm BL}}=2 g_{\rm BL} v_{\rm BL}, M_{N_{i}}=\sqrt{2}Y_{N_{i}}v_{\rm BL}$ where $v$ and $v_{BL}$ are the vacuum expectation values (VEVs) of $H$ and $\Phi$ respectively. \\



\noindent {\bf Miracle-less WIMP Dark Matter:}
For $B-L$ gauge symmetry breaking around the TeV corner and $g_{\rm BL}\sim \mathcal{O}(1)$, DM can freeze-out non-relativistically, falling into the standard WIMP paradigm, as long as DM mass remains below a few hundred TeV \cite{Griest:1989wd}. We consider high scale symmetry breaking namely, large $v_{\rm BL}(M_{Z_{\rm BL}})$ to ensure that DM freezes out at an early epoch while it is still relativistic. 

\begin{figure}[h]
\includegraphics[scale=0.5]{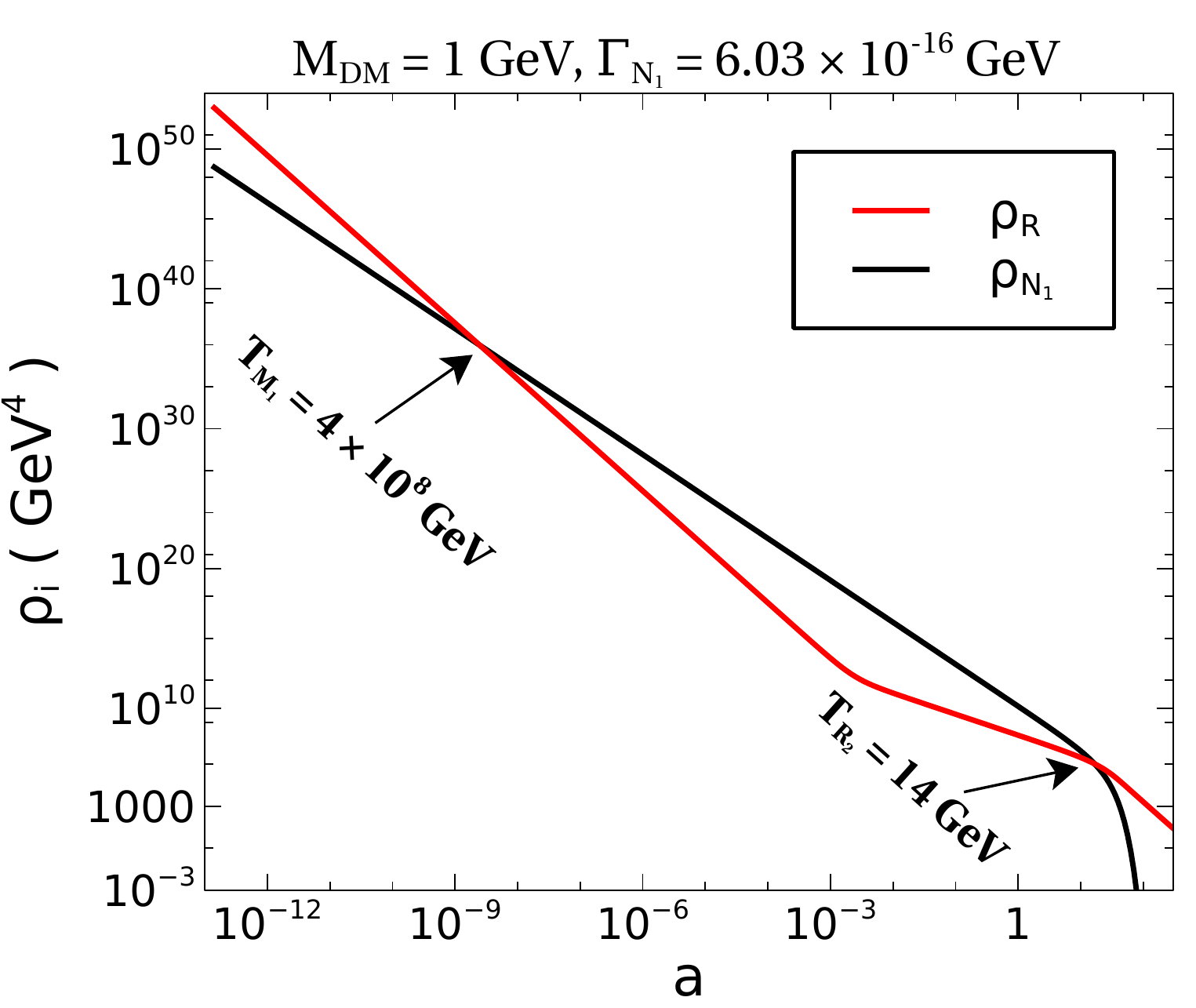}\\
\caption{The evolution of radiation (red) and $N_1$ (black) densities for $M_{\rm DM}=1$ GeV, $M_{N_1}=7\times 10^{10}$ GeV, $v_{\rm BL}=2.5\times 10^{12}$ GeV, $g_{\rm BL}=0.2$, $\Gamma_{N_1}=6.03\times 10^{-16}$ GeV such that $\Omega_{\rm DM} h^2\sim 0.12$. Here, the scale factor is normalised such that $a_{\rm initial}=\left(\frac{1 {\rm ~GeV}}{T_{\rm initial}}\right)$ with $T_{\rm initial}(\simeq v_{\rm BL})$ is considered to be the initial temperature of the initial radiation dominated universe.}
\label{fig:BPcombined} 
\end{figure}

Now, a large $M_{Z_{\rm BL}}$ implies weaker annihilation rate of the DM to visible sector particles and hence generally leads to overabundance provided the DM is thermalised at early universe. This overabundance can be brought down by the late decay of one of the RHNs say, $N_1$ after the DM freeze-out, which injects entropy ($s$) into the thermal bath \cite{Scherrer:1984fd}. In order to realise this possibility of sizeable entropy production, it is necessary for the long-lived $N_1$ to dominate the energy density of the universe at late epochs. The key phases in the universe relevant for our discussion are summarised in Fig.\,\ref{fig:Sketch} showing the intermediate matter domination phase MD1 due to long-lived $N_1$ \cite{Allahverdi:2021grt}. A precise description of such non-standard universe is possible by considering a system of coupled Boltzmann equations which govern the evolutions for the temperature of the universe ($T$) and comoving number densities of both DM and the diluter $N_1$ \cite{Nemevsek:2012cd, Bezrukov:2009th, Borah:2017hgt, Cirelli:2018iax, Dror:2020jzy,Dutra:2021lto, Arcadi:2021doo, Borah:2021inn}. The Fig.\,\ref{fig:BPcombined} shows the evolution of radiation and $N_1$ densities for benchmark parameters where we observe an intermediate phase of $N_1$ domination. The standard radiation dominated phase RD2 is recovered after $N_1$ decays with the Hubble parameter $H\propto \frac{T^2}{M_P}$.

The DM relic density, for relativistic freeze-out, is given by \cite{Kolb:1990vq}
\begin{align}
 \Omega_\chi h^2 & =2.745\times 10^{8}\times Y_\infty m_\chi,\label{eq:mainEQ}
\end{align}
where $Y_{\infty}=\frac{0.278}{g_{*s}(x_f)}\times \frac{3 g_\chi}{4}$ with $g_{\chi}$ and $g_{*s}(x_f)$ are the DM internal dof and entropy dof of the Universe at DM freeze-out temperature respectively. We also consider $g_{*s}(x_f)=106.75$ for SM dof. If $\chi$ leads to overabundance, the required entropy dilution factor $S = \Omega_\chi h^2/0.12$ can be approximated as \cite{Scherrer:1984fd},
\begin{align}
 S\simeq \left[2.95\times \left(\frac{2\pi^2 \tilde{g}_{*}(T_{N_1})}{45}\right)^{1/3}\frac{(r M_{N_1})^{4/3}}{(\Gamma_{N_1} M_P)^{2/3}}\right]^{3/4},\label{eq:dilutionF}
\end{align}
where $\tilde{g}_{*}(T_{N_1})$ is the number of relativistic dof during $N_1$ decay at $T=T_{N_1}$. The parameter $r$ is the freeze-out number density of $N_1$. 
Assuming instantaneous decay of $N_1$ and considering relativistic freeze out for $N_1$ we find,
 \begin{align}
  T_{N_1}\simeq 3.104\times 10^{-10}\left(\frac{M_{N_1}}{m_\chi}\right) \rm GeV.\label{eq:anaEntropy}
 \end{align}
 The temperature $T_{N_1}$ is inversely proportional to DM mass and approximately marks the end of early matter domination in the assumption of instantaneous $N_1$ decay. In principle, $T_{N_1} $ should be nearly equal to $T_{R_2}$. The obtained $T_{N_1}$ can be easily translated to $\Gamma_{N_1}$ using $H(T_{N_1})=\Gamma_{N_1}$ which is the required decay width to bring down the relic within observed limit. For the benchmark point in Fig.\,\ref{fig:BPcombined}, we found $T_{N_1}=21.72$ GeV from Eq. (\ref{eq:anaEntropy}). In the numerical analysis for the same benchmark point we earlier obtained $T_{R_2}=14$ GeV. The small discrepancy between numerical and analytical estimate emerges due to the approximation of instantaneous decay of $N_1$ in the analytical computation.
 
\noindent {\bf Gravitational Wave Spectral Shape with Miracle-less WIMP:}
Cosmic strings\cite{Kibble:1976sj, Nielsen:1973cs}, one of the potential sources of primordial GW, have gained a great deal of attention after the recent finding of a stochastic common spectrum process across many pulsars\cite{NANOGrav:2020bcs,Goncharov:2021oub,Ellis:2020ena,Blasi:2020mfx,Samanta:2020cdk}. CS appear as topological defects after spontaneous breaking of a symmetry group containing a vacuum manifold which is not simply connected\cite{Nielsen:1973cs}. The simplest group that exhibits such feature is $U(1)$ which naturally appears in many theories beyond the SM. Numerical simulations\cite{Ringeval:2005kr,Blanco-Pillado:2011egf} based on Nambu-Goto action indicate that dominant energy loss from a string loop is in the form of GW radiation if the underlying symmetry is gauged. The resulting GW background is detectable when the symmetry breaking scale $\Lambda_{CS}\gtrsim 10^9$ GeV--a fact that makes CS an outstanding probe of super-high scale physics\cite{Buchmuller:2013lra, Dror:2019syi, Buchmuller:2019gfy,King:2020hyd, Fornal:2020esl,  Buchmuller:2021mbb, Masoud:2021prr, Afzal:2022vjx}. This includes the present scenario of Miracle-less WIMP featuring a high-scale ($\Lambda_{\rm CS}\equiv v_{\rm BL}$) breaking of a gauged $U(1)_{B-L}$. The properties of CS are described by their normalised tension $G \mu\sim G \Lambda_{\rm CS}^2$ with $G$ being the Newton's constant. Unless the motion of a long-string network gets damped by thermal friction\cite{Vilenkin:1991zk}, shortly after formation, the network oscillates ($t_{\rm osc}$) and  enters the scaling regime\cite{Blanco-Pillado:2011egf,Bennett:1987vf,Bennett:1989ak} which is an attractor solution of two competing dynamics--stretching of the long-string correlation length due to cosmic expansion and fragmentation of the long strings into loops which oscillate to produce particle radiation or GW\cite{Vilenkin:1981bx,Turok:1984cn,Vachaspati:1984gt}. 

A set of normal-mode oscillations with frequencies $f_k=2k/l$ constitute the total energy loss from a loop, where the mode numbers $k=1,2,3....\infty$. Therefore, the GW energy density parameter is defined as $\Omega_{\rm GW}(t_0,f)=\sum_k\Omega_{\rm GW}^{(k)}(t_0,f)$, with $t_0$ being the present time and $f\equiv f(t_0)= f_k a(t_0)/a(t)$. Present day GW energy density corresponding to the mode $k$ is computed with the integral \cite{Blanco-Pillado:2013qja} 
\begin{align}
 \Omega_{\rm GW}^{(k)}(t_0,f)=\frac{2k G\mu^2\Gamma_k}{f \rho_c}\int_{t_{osc}}^{t_0}dt\left[\frac{a(t)}{a(t_0)}\right]^5 n\left(t,l_k\right),\label{gwint}
\end{align}
where $n\left(t,l_k\right)$ is a scaling loop number density which can be computed analytically using Velocity-dependent-One-Scale (VOS) \cite{Martins:1996jp,Martins:2000cs,Auclair:2019wcv} model \footnote{Compared to the numerical simulation, VOS model overestimates loop number density by a factor of 10. We therefore use a normalization factor $\mathcal{F}_\alpha=0.1$ to be consistent with simulation \cite{Auclair:2019wcv}.}, $\rho_c$ is the critical energy density of the universe and $\Gamma_k=\frac{\Gamma k^{-\delta}}{\zeta(\delta)}$ depends on the small scale structures in the loops such as cusps ($\delta=4/3$) and kinks ($\delta=5/3$). In this article, we consider only cusps to compute GW spectrum.
\begin{figure}[h]
\includegraphics[scale=0.45]{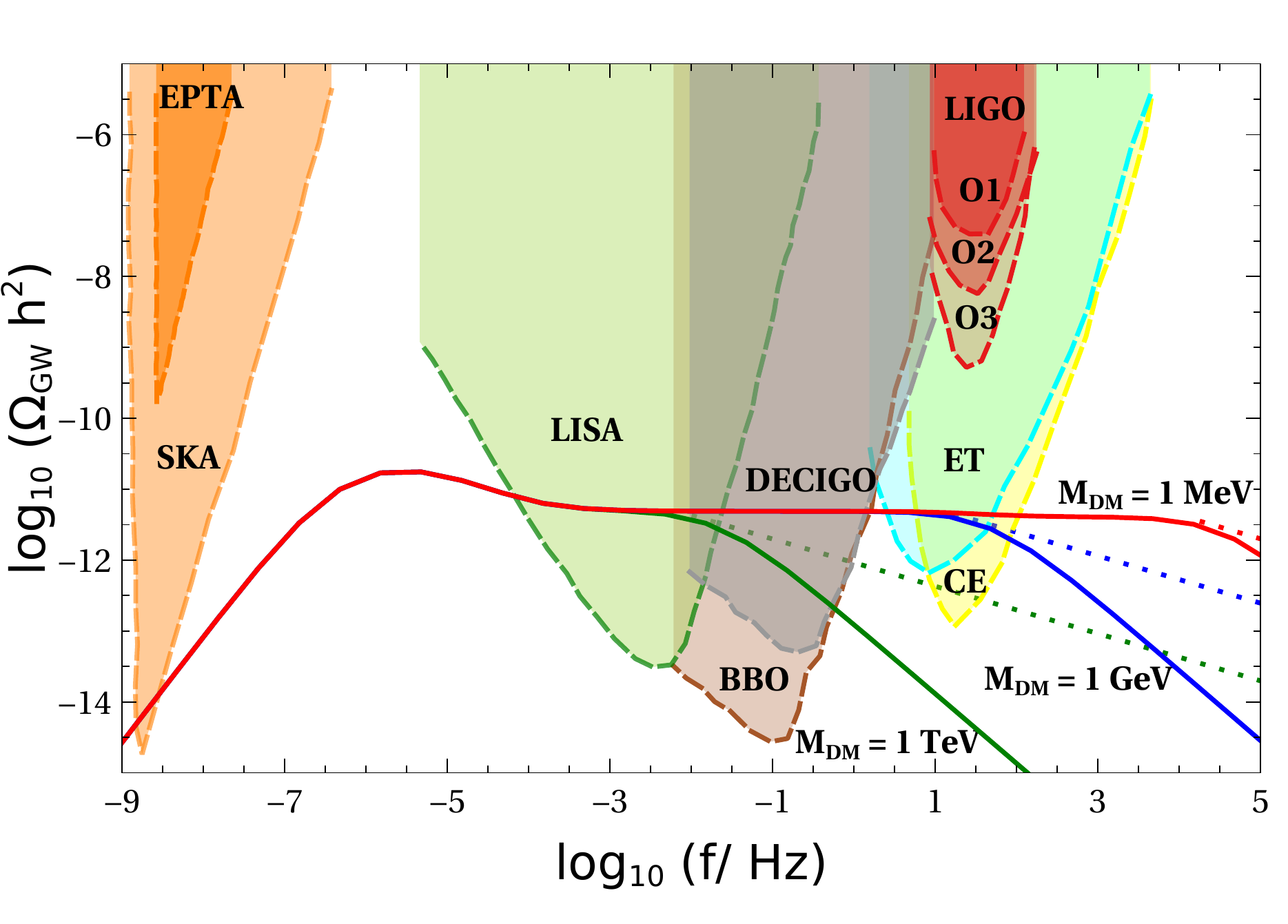}
\caption{Fundamental mode ($k=1$) GW spectra for the benchmark point in Fig.\,\ref{fig:BPcombined} with different combinations of $M_{\rm DM}, \Gamma_{N_1}$ to satisfy correct relic. The dotted lines indicate the spectral behaviour beyond the turning point frequency when all the modes are summed.}
\label{fig:GW_cs}  
\end{figure}  
A typical feature of GWs from CS is a flat plateau (e.g., the red solid line ($f\gtrsim 10^{-3}$ Hz in Fig.\,\ref{fig:GW_cs}) due to loop formation and decay during radiation domination, with an amplitude given by 
\begin{align}
\Omega_{\rm GW}^{(k=1),{\rm plateau}}(f)=\frac{128\pi G\mu}{9\zeta(\delta)}\frac{A_r}{\epsilon_r}\Omega_r\left[(1+\epsilon_r)^{3/2}-1\right], \label{flp1}
\end{align}
where $\epsilon_r=\alpha/\Gamma G\mu$ with $\alpha$ the initial (at $t=t_i$) loop size parameter, $\Omega_r\simeq 9\times 10^{-5}$ and $A_r=5.4$ \cite{Auclair:2019wcv}. In our analysis, we have considered $\alpha \simeq 0.1$\,\cite{Blanco-Pillado:2013qja,Blanco-Pillado:2017oxo} (as suggested by simulations),  $\Gamma\simeq 50$\,\cite{Vachaspati:1984gt}, and CMB constraint $G\mu\lesssim 10^{-7}$ \cite{Charnock:2016nzm} which lead to $\alpha \gg \Gamma G \mu$. In this limit, Eq.\eqref{flp1} implies $\Omega_{\rm GW}^{(k=1)}(f)\sim \Lambda_{\rm CS}$, a property that makes models with larger breaking scales more testable with GWs from CS. Interestingly, if there is a matter domination which in our case is provided by the long-lived $N_1$, the plateau breaks\cite{Cui:2017ufi,Cui:2018rwi,Gouttenoire:2019kij} at a turning point frequency $f_\Delta=\sqrt{\frac{8}{\alpha\Gamma G\mu}}t_\Delta^{-1/2}t_0^{-2/3}t_{\rm eq}^{1/6}$, where $t_\Delta$ and $t_{\rm eq}$ are the times at the end of the early matter domination at a temperature $T_\Delta\equiv T_{N_1}\simeq T_{R_2}$ and at standard matter radiation equality at a temperature $T_{M_2}$ respectively. Beyond $f_\Delta$, the spectrum goes as $\Omega_{\rm GW}\sim f^{-1}$ for $k=1$ mode (when infinite modes are summed, $\Omega_{\rm GW}\sim f^{-1/3}$\cite{Blasi:2020wpy,Gouttenoire:2019kij,Datta:2020bht, Gouttenoire:2019rtn}). This spectral behaviour therefore, serves as a probe of an early matter domination which in our case is a natural requirement to obtain correct relic density of Miracle-less WIMPs.

\begin{figure*}[t]
\includegraphics[scale=0.5]{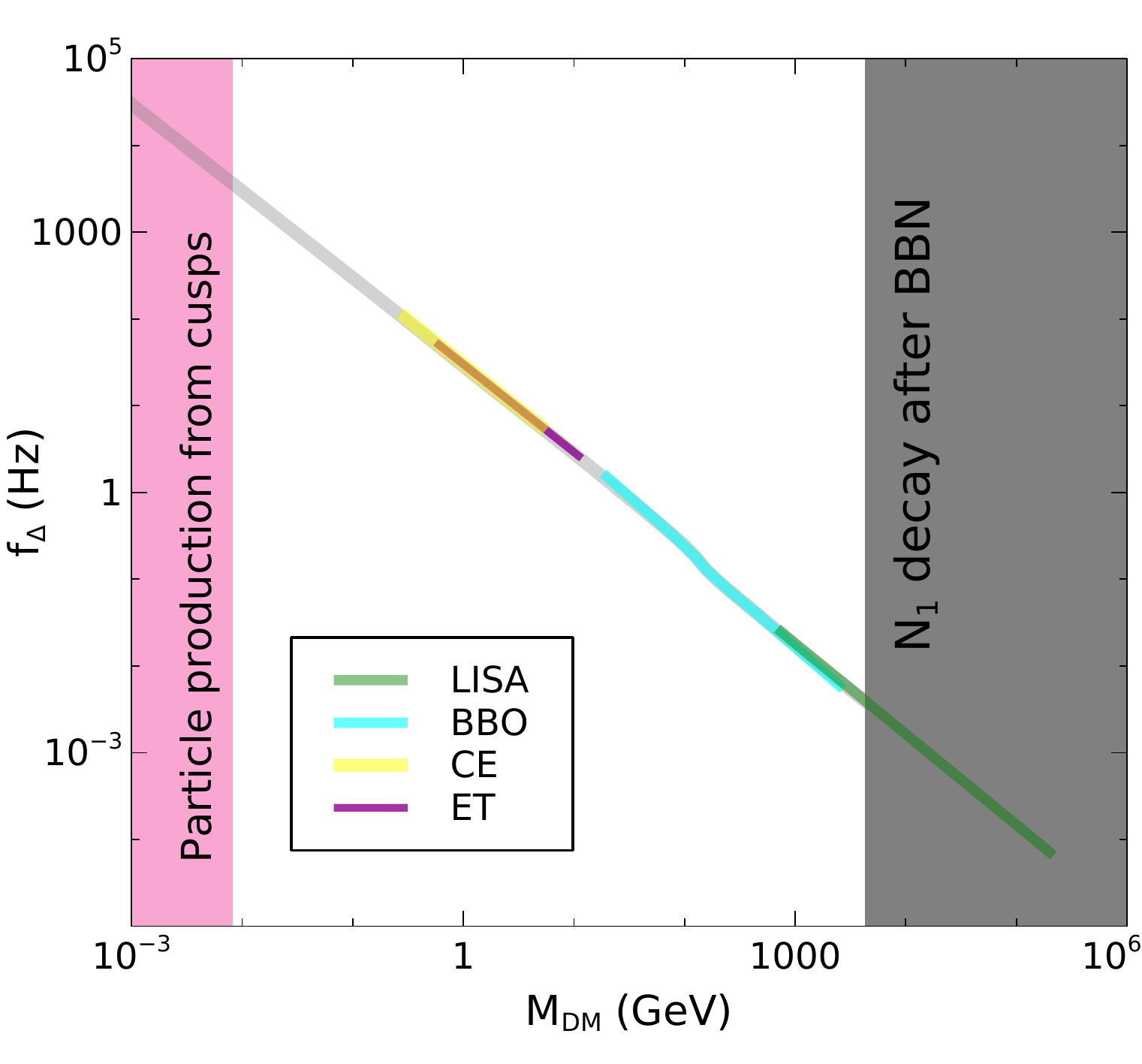}
\includegraphics[scale=0.5]{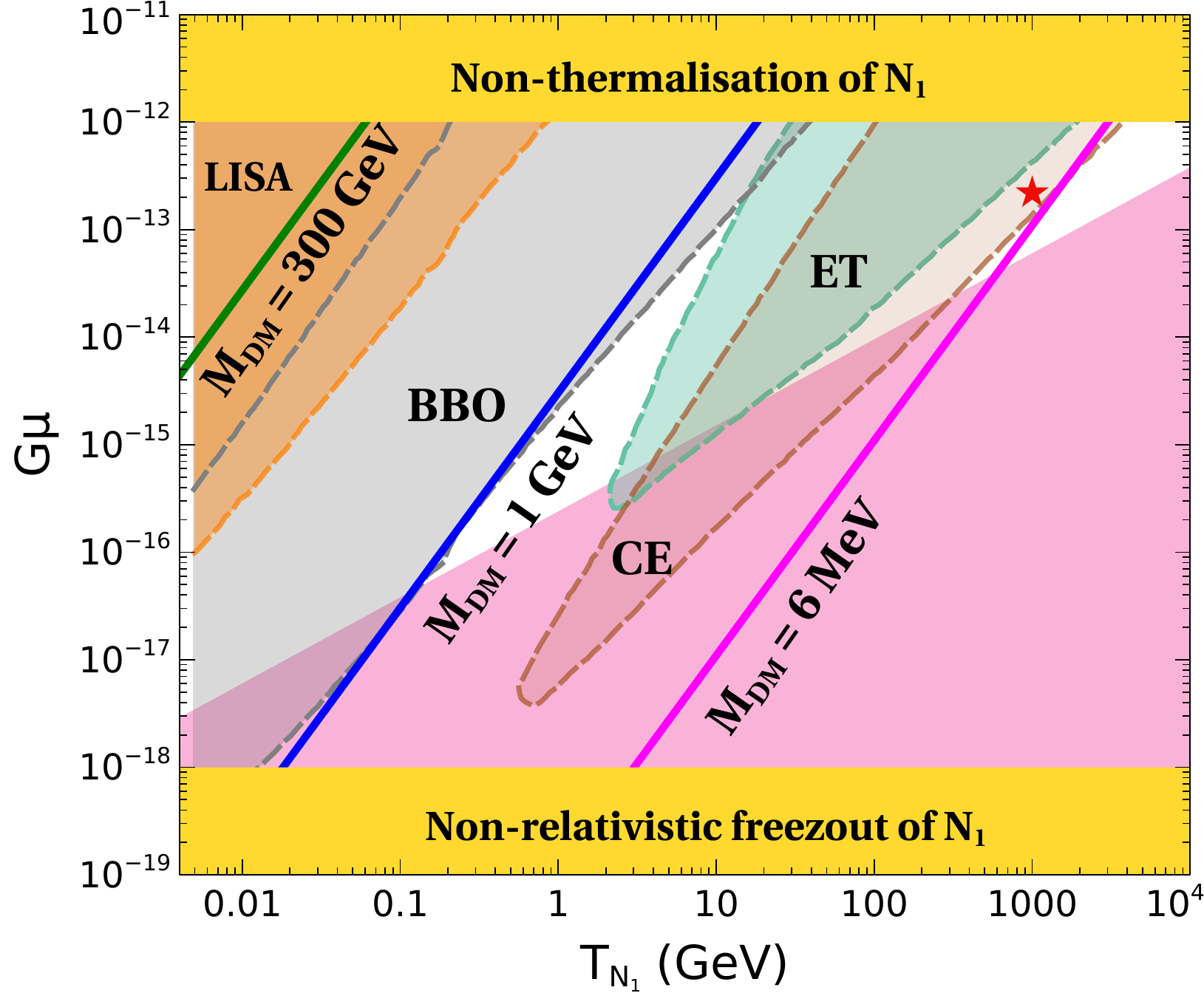}
\caption{Left: Contour representing DM relic $\Omega_{\chi} h^2=0.12$ in the $f_{\Delta}-M_{\rm DM}$ plane for fixed $v_{\rm BL},{g_{\rm BL}}$ and $M_{N_1}$ values (as in Fig.\,\ref{fig:BPcombined}) along with the future sensitivities of different GW detectors. Right: Contours in the $T_{N_1}-G\mu$ plane for three different DM masses satisfying correct relic, along with future GW detector sensitivities for a chosen benchmark point $g_{\rm BL}=0.2$ and $Y_{N_1}=0.03 g_{\rm BL}$. The symbol $`\star$' indicates a benchmark point satisfying leptogenesis requirements. In the pink region of both the figures, the particle production dominates over the GW emission and hence discarded.}
\label{fig:fDelta-mDM} 
\end{figure*} 

Similar turning points ($f_c$) can be obtained from the lower bound on $t_i>t_c$ that corresponds to $\alpha t_i>l_c$ \cite{Gouttenoire:2019kij}, where $l_c$ is a critical length below which particle production becomes dominant. Therefore, to claim the spectral break to be a consequence of the end of a matter era, we should have $f_\Delta < f_c$. For cusps like structures, $l_c=\frac{\mu^{-1/2}}{(\Gamma G \mu)^2}$ \cite{Auclair:2019jip} which translates to a lower bound on $G\mu$ as
\begin{align}
    G\mu\gtrsim 2.4\times 10^{-16} T_{\Delta}^{4/5}.
\end{align}
Let us also mention that the temperature $T_c$ corresponding to $l_c\equiv \alpha t_c$ always lies between $T_{M_1}$ and $T_{R_2}$. This means the loops which effectively contribute to the GWs, are formed in the $N_1$ dominated epoch, making the very first radiation domination (RD1) irrelevant to the computation.
 
We compute the $\Omega_{\rm GW}$ for the three DM masses, differing by order of magnitudes, as shown in Fig.\,\ref{fig:GW_cs}. We tune $\Gamma_{N_1}$ appropriately for each of the DM masses such that $\Omega_{\chi} h^2=0.12$.
The relevant model parameters other than $M_{\rm DM}$ and $\Gamma_{N_1}$ are considered to be the same as in Fig.\,\ref{fig:BPcombined} which precisely determine the $G\mu$=$1.32\times 10^{-13}$. Since $M_{N_1}$ and $g_{\rm BL}$ are fixed, the MD1 phase starts at the same temperature for all the three DM masses in Fig.\,\ref{fig:GW_cs}. 
Within such non-standard framework, the GW spectra is expected to show spectral break ($f_\Delta$) at higher frequency depending on the value of $T_{N_1}$. In general, a lower $T_{N_1}$ implies a longer period of MD1 phase and further leads to a smaller spectral break frequency $f_\Delta$. Previously in Eq.\,(\ref{eq:anaEntropy}) we have derived an unique relationship between $T_{N_1}$ and DM mass where $T_{N_1}\propto \frac{1}{M_{\rm DM}}$. This leads to a one-to-one correlation between $f_\Delta$ and DM mass $M_{\rm DM}$ with a larger DM mass resulting in smaller $f_\Delta$ as observed in Fig.\,\ref{fig:GW_cs}. This feature is also prominent in left panel of Fig.\,\ref{fig:fDelta-mDM}, where we have determined the $f_\Delta$ for different order of DM mass ranging from 1\,MeV to 1\,PeV.
We also mark different parts of the $M_{\rm DM}-f_\Delta$ line following various planned experimental sensitivities. The region which escapes the detection sensitivities is highlighted in light grey, corresponding to lighter DM. The success of the big bang nucleosynthesis (BBN) restricts the DM mass to remain below 8 TeV while for $M_{\rm DM}\lesssim 10$\,MeV, particle production from cusps dominates and hence disfavored. We find that a substantial range of the DM mass (100 MeV-8 TeV), allowed by relevant constraints, is within the reach of GW detectors like LISA\,\cite{LISA:2017pwj}, BBO\,\cite{Yagi:2011wg}, ET\,\cite{Punturo_2010}, CE\,\cite{LIGOScientific:2016wof} keeping the Miracle-less WIMP verifiable or falsifiable in near future.

In the right panel of Fig.\,\ref{fig:fDelta-mDM}, we obtain the predictions for three different DM masses in the $T_{N_1}$-$G\mu$ plane such that the relic bound is satisfied. We have used Eq.(\ref{eq:anaEntropy}) with the assumption $T_{R_2}\simeq T_{N_1}$ and expressed $M_{N_1}$ as function of $G{\mu}$ by fixing $g_{\rm BL}=0.2$ and $Y_{N_1}=0.03g_{\rm BL}$. The future sensitivities of different experiments namely LISA\,\cite{LISA:2017pwj}, BBO\,\cite{Yagi:2011wg}, ET\,\cite{Punturo_2010} and CE\,\cite{LIGOScientific:2016wof} are also assembled in the same plane. For any particular point with specific ($T_{N_1},G\mu$) coordinates along a fixed $m_{\rm DM}$ line, one can compute the turning frequency $f_\Delta$. This figure illustrates whether any of the above-mentioned future GW experiments has the ability to probe that particular $f_\Delta$. For DM mass of 300 GeV, the corresponding $f_\Delta$ falls within the LISA sensitivity\,\cite{LISA:2017pwj}. On the other hand, the proposed sensitivity of CE \cite{LIGOScientific:2016wof} experiment can probe DM mass as light as 6 MeV in the present framework. 
We also point out the region in $T_{N_1}-G\mu$ plane where thermalisation of $N_1$ can not be achieved in early universe (yellow shaded region on top). The yellow shaded bottom region favors non-relativistic freeze-out for $N_1$ and hence our analytical derivation of $T_{N_1}$ in Eq. (\ref{eq:anaEntropy}) does not remain valid. 

\noindent\textbf{Leptogenesis with Miracle-less WIMP:} The presence of RHNs in the $B-L$ model can also explain the origin of baryon asymmetry of the universe \cite{Iso:2010mv} via leptogenesis \cite{Fukugita:1986hr}. Since $N_1$ being long-lived has tiny Yukawa couplings, we consider lepton asymmetry to be generated from two heavier RHNs. Due to significant entropy dilution at late epoch, the resonant leptogenesis mechanism \cite{Pilaftsis:2003gt} appears to be the suitable one to achieve the observed $\eta_B$. The present day baryon to photon ratio is conventionally parametrised as a function of lepton asymmetry parameter ($\varepsilon$) and efficiency factor ($\kappa_f$) as \cite{Buchmuller:2004nz},
\begin{align}
\eta_B=\frac{3}{4}\frac{a_{\rm sph}}{f} \sum_i \varepsilon_i\kappa_f^i.
\end{align}
where $a_{\rm sph} = \frac{28}{79}$ is the fraction of $B-L$ asymmetry converted into a baryon asymmetry by sphaleron processes, and $f=\frac{2387}{86}$ is the dilution factor. 

We have followed the the Casas-Ibarra (CI) parametrization \cite{Casas:2001sr} to express the Dirac neutrino mass matrix in terms of the leptonic mixing matrix $U_{\rm PMNS}$, the light and heavy neutrino masses and a complex orthogonal matrix $\mathcal{R}$ as \begin{equation}
Y_D = \frac{1}{v}{\sqrt{M_N}} R \sqrt{\hat{m_\nu}} U^\dagger_{\rm PMNS} 
\end{equation}
with $\hat{m_\nu} \equiv \textrm{Diag}(m_1,m_2,m_3)$ and $M_N$ being the diagonal light and heavy neutrino mass matrices respectively. For three right handed neutrinos taking
part in seesaw mechanism, $\mathcal{R}$ is function of three complex rotation parameters. Assuming one of them (rotation in 1–2 sector) to be vanishing, we can write
\begin{align}
    \mathcal{R}=
    \left(
\begin{array}{ccc}
 \cos \delta^\prime & 0 & \sin \delta^\prime \\
 -\sin \delta \sin \delta^\prime & \cos \delta & \sin \delta \cos \delta^\prime \\
 -\cos \delta \sin \delta^\prime & -\sin \delta & \cos\delta \cos \delta^\prime \\
\end{array}
\right)
\end{align}
Since $N_1$ is long-lived with very small decay width in our set-up
making it effectively decoupled from seesaw mechanism, we can approximately set $\delta^\prime\sim 0$. In this limit, effectively the Dirac Yukawa coupling $Y_D$ represents a 2 × 3 matrix in flavour basis.
The lepton asymmetry can be generated from the decays $N_2$ and $N_3$. In order to obtain resonant enhancement to the lepton asymmetry parameter, $N_2$ and $N_3$ need to be nearly degenerate with their mass splitting expressed as $\frac{M_{N_3}-M_{N_2}}{M_{N_2}}=\Delta$. 
As a benchmark point, we make the following choices of the relevant parameters,
\begin{align}
M_{N_2}=10^{11}\,{\rm GeV}, ~\Delta=2.6\times 10^{-6},~ \delta=0.71+0.42 i.  \nonumber
\end{align}
The mass scale of $N_2$ is chosen to be high such that it decays completely during the very first radiation dominated epoch. Considering the lightest neutrino mass to be vanishing, we have used the best fit values of experimentally observed neutrino oscillation parameters. Correspondingly we obtain $\varepsilon_{2,3}=0.33$ and $\kappa_f^{2}\sim 7.8\times 10^{-3}$, $\kappa_f^{3}\sim 6.7\times 10^{-3}$ using the standard expressions as noted in \cite{Pilaftsis:2003gt}. These choices lead to $\eta_B\simeq 4.58\times 10^{-5}$ in the standard scenario which is clearly overproduced. Therefore, to match with the present day observed value for $\eta_B$, the required amount of entropy dilution factor is $7.5\times 10^{4}$ which corresponds to the DM mass $\sim$ 8.5 MeV. For $T_{N_1}\sim 1$\,TeV, the 8.5 MeV DM mass implies $G\mu\simeq 2\times 10^{-13}$ with $Y_{N_1}=0.03g_{\rm BL}$ and $g_{\rm BL}=0.2$ (from Eq.(\ref{eq:anaEntropy})) as marked by red colored $`\star'$ symbol in right panel plot of Fig.\,\ref{fig:fDelta-mDM}. This particular point in the $T_{N_1}-G\mu $ plane which represents the simultaneous satisfaction of DM relic abundance and baryon asymmetry of the universe is verifiable by CE experiment.
A more rigorous analysis including lepton flavour effects \cite{Abada:2006fw, Abada:2006ea, Nardi:2006fx, Blanchet:2006be} is left for future studies. \\


\noindent \textbf{Conclusion:} We have proposed a novel way of probing DM having a wide range of masses by observations of spectral breaks in GW spectrum generated by cosmic strings, with complementary predictions which distinguish it from a general non-standard cosmological scenario with similar impacts on GW spectrum. Presence of a high scale Abelian gauge symmetry breaking results in generation of cosmic strings with observable GW spectrum while also causing insufficient DM annihilations mediated by superheavy gauge boson. Thermally generated DM with such insufficient annihilation rates, dubbed as Miracle-less WIMP, leads to overproduced relic abundance, requiring late entropy dilution. Adopting the minimal gauged B-L framework with several other motivations, we ensure such entropy dilution due to late decay of one of the RHNs. Depending upon DM mass, one requires different decay time of such diluter, leading to unique turning frequencies in the usually flat GW spectrum generated by cosmic strings. For a wide range of DM mass, such turning frequencies remain within reach of next generation GW experiments like LISA, BBO, CE, ET. The heavier two RHNs generate light neutrino masses as well as the baryon asymmetry of the universe via leptogenesis. We have observed that simultaneous realisation of successful leptogenesis and DM relic is possible in the present framework with the turning frequency falling within reach of the above GW experiments. The model also predicts vanishingly small lightest active neutrino mass $m_{\rm lightest} \leq 10^{-20}$ eV. While this will keep the effective neutrino mass much out of reach from ongoing tritium beta decay experiments like KATRIN \cite{KATRIN:2019yun}, future experiments should be able to confirm or refute it. Additionally, near future observation of neutrinoless double beta decay (NDBD) \cite{Dolinski:2019nrj} can also falsify our scenario, particularly for normal ordering of light neutrinos. This is due to the fact that future NDBD experiments can probe normal ordering only for $m_{\rm lightest} > 10^{-2}$ eV which lies much above the tiny value predicted in our scenario. \\

\noindent \textbf{Acknowledgements:}~AKS is supported by
NPDF grant PDF/2020/000797 from Science and Engineering Research Board, Government of India. RS is supported by the MSCA-IF IV FZU $- CZ.02.2.69/0.0/0.0/20 079/0017754$ project and acknowledges European Structural and Investment Fund and the Czech Ministry of Education, Youth and Sports.

\twocolumngrid

\end{document}